\documentstyle[prb,aps,twocolumn,epsfig]{revtex}

\begin{document}
\title{Raman Scattering cross section of Spin Ladders}
\draft
\author{E. Orignac$^{a,b}$ and R. Citro$^{c}$}
\address{$^{a}$Serin Physics Lab, Rutgers University, Piscataway NJ
08855-0849 \\
$^{b}$ LPTENS, 24 Rue Lhomond, 75231 Paris Cedex 05 \\
$^{c}$Dipartimento di Scienze Fisiche ''E.R. Caianiello'',\\
University of Salerno \\
and Unit\`{a} INFM of Salerno, Baronissi\\
(Sa), Italy}
\wideabs{
\date{\today}
\maketitle

\begin{abstract}
The Raman scattering spectra from magnetic excitations in an
antiferromagnetic spin-$\frac 1 2$
 two leg ladder is investigated for weak and strong
interladder coupling. In the first case, a cusp in the Raman intensity
is obtained at a frequency twice the gap. In the second case, a peak
at twice the gap  replaces the cusp. We discuss the relevance of our
calculation to recent experiments on $\mathrm{CaV_{2}O_{5}}$ and
$\mathrm{Sr_{14}Cu_{24}O_{41}}$.
\end{abstract}
\pacs{PACS: 75.40.Gb 78.30-j 75.10.Jm }
}

Raman scattering is an experimental technique that has provided 
valuable informations about the spin dynamics in quasi-one dimensional
antiferromagnets in the recent years. Since Raman scattering is
sensitive to singlet excitations, this 
technique is complementary of neutron diffraction which is sensitive
to triplet excitations.  It has been used to probe spin
1/2 chains \cite{yamada_raman_1d}, spin 1 chains
\cite{sulewski_raman_spin1}, spin Peierls systems \cite{gugeo3_lemmens,loosdrecht_raman_cugeo3,kuroe_raman_nav2o5,golubchik_raman_nav2o5}, and spin
ladders
\cite{srcuo_raman_spectra,cavo_raman_spectra,sugai_raman_sr14cu24o41}.
In particular, Raman scattering has been very useful in the analysis
of magnetic excitations in the spin-Peierls compound
$\mathrm{CuGeO_3}$. The bosonized theory of 
dimerized spin $1/2$ states that is believed to describe the dimerized
low temperature phase of spin-Peierls systems 
predicts the appearance of a singlet bound
state of two triplet excitations 
at an energy $\sqrt{3}\Delta$, where $\Delta$ is the spin
gap\cite{uhrig_dimerized}. Such singlet bound state has been
successfully observed in Raman scattering experiments on ${\rm
CuGeO_3}$ for $T<T_{\text{SP}}$ 
at an energy $1.79\Delta$, close to the theoretical
prediction. Moreover, the peak was not observed in the uniform
phase($T>T_{\text{SP}}$),
showing that it is characteristic of the dimerized
phase\cite{loosdrecht_raman_cugeo3,lemmens_dimerized}. For
$T>T_{\text{SP}}$, a broad band of magnetic excitations is
observed\cite{loosdrecht_raman_cugeo3}.  
 The theoretical
analysis of magnetic Raman scattering is based on the Fleury-Loudon
Hamiltonian\cite{fleury_raman,moriya_raman}, that describes the interaction of
photons with magnetic excitations.
 There exists at present a
certain amount of litterature on the theory of Raman scattering from
dimerized spin chains, both analytical
\cite{uhrig_dimerized,brenig_raman_frustrated,bouzerar_breathers} and numerical
\cite{augier}. The case of frustrated spin chains has also been
investigated\cite{muthukumar_raman_frustrated}, in relation with the
Raman spectra of $\mathrm{CuGeO_3}$ at $T>T_{\text{SP}}$. The theory
of Ref. \onlinecite{muthukumar_raman_frustrated} reproduced well the
features of the spectrum at $T>T_{\text{SP}}$ \cite{loosdrecht_raman_cugeo3}. 
The application of Raman scattering to probe the singlet excitations
of two leg ladders is more
recent\cite{srcuo_raman_spectra,cavo_raman_spectra,sugai_raman_sr14cu24o41}. 
In spin ladder systems, magnetic peaks in the Raman intensity were observed at
twice the spin gap. 
From the theoretical point of view,
  some numerical calculations are available\cite{spinladder_raman_numerics},
but no analytic expression of the Raman intensity has been derived so
far. In order to fill this gap,  we discuss in the present work the
Raman spectrum of an 
antiferromagnetic spin-1/2 ladder.
After recalling some basic results on the Fleury-Loudon theory of
magnetic Raman scattering, 
we will consider first  the Majorana fermion approach valid for weak
coupling and then the Bond Operator Technique (BOT) valid for 
the strong coupling
case. The Majorana fermions approach leads to a cusp in the Raman
intensity at twice the gap, in disagreement with experiment. 
We discuss briefly what
could be missing in the Majorana fermions description. On the other
hand, the BOT predicts correctly the presence of peaks in the Raman
intensity at twice the gap.

We consider two coupled antiferromagnetic $S=1/2$ Heisenberg chains, whose
Hamiltonian is

\begin{eqnarray}
H=J\sum_{i}\left( \overrightarrow{S}_{1,i}\overrightarrow{S}%
_{1,i+1}+\overrightarrow{S}_{2,i}\overrightarrow{S}%
_{2,i+1}\right) \nonumber \\ +J_{\perp }\sum_{i}\stackrel{\rightarrow
}{S}_{1,i} \cdot \stackrel{%
\rightarrow }{S_{2,i}}  \label{hladder}
\end{eqnarray}
where $J>0$ and $J_{\perp }>0$ denotes the intra- and inter- chain
antiferromagnetic interactions, respectively. The interaction of light
with the antiferromagnetic fluctuations is described by  Loudon-Fleury's 
\cite{fleury_raman,moriya_raman} photon-induced super-exchange operator

\begin{equation}\label{eq:loudon_fleury}
H_{R}=\sum_{i,j}(\overrightarrow{{E}_{I}}\cdot \overrightarrow{%
\delta }_{ij})(\overrightarrow{{E}_{S}}\cdot \overrightarrow{%
\delta }_{ij})\overrightarrow{{S}_{i}}\cdot \overrightarrow{S_{j}}
\label{hr}
\end{equation}
where $\overrightarrow{E}_{I}(\overrightarrow{E}_{S})$ are the incident (scattered) electric field,
and $\overrightarrow{\delta }_{ij}$ is a unit vector connecting 
the sites $i$ and $j$, at which the spins $\overrightarrow{S}_{i}$%
and $\overrightarrow{S}_{j}$ are located. A derivation of (\ref{hr})
starting from the Hubbard Hamiltonian can be found in
Ref. \cite{shastry_raman}. 

The Raman cross section\cite{loudon_raman,reiter_raman}
 can be expressed as a function of the
retardated Raman response function as:
\begin{equation}
\frac{d^{2}\sigma }{d\Omega d\omega _{2}}=\frac{\omega_1
\omega_2^3}{2\pi c^4 V} \frac {n_2}{n_1} \frac 1 {1-e^{-\beta \hbar
\omega}} { \rm Im} \chi_R(\omega)
\end{equation}
 $\omega _{1}$ and $\omega _{2}$ are the frequencies of the incoming
and scattered radiation, respectively, $\omega=\omega_2-\omega_1$,
$n_1$ and $n_2$ are the respective refractive index. $V$ is the volume
of the crystal and $c$ the velocity of light. 
 The retardated linear response function $\chi_R(\omega)$ is defined
as:
\begin{equation}
\chi _{R}^{ret}(\omega)=\frac{i}{\hbar}\int_0^{\infty}
e^{i(\omega+i 0) t} { \rm Tr}\left\{ Z^{-1}e^{-\beta H}\left[ H_{R}(t),H_{R}(0)\right]
\right\} ,  \label{ramansusc}
\end{equation}
 where $Z={ \rm Tr}e^{-\beta H}$ and  $H_{R}$ is the Loudon-Fleury
Hamiltonian (\ref{eq:loudon_fleury}).

By inserting the resolution of identity in (\ref{ramansusc}), the Raman
intensity can be written as

\begin{eqnarray}
\frac{d^{2}\sigma }{d\Omega d\omega _{2}} \propto \frac{1}\hbar\frac 1
Z \sum\limits_{n,m}e^{-\beta E_{n}}\left| \left\langle \Psi _{n}\left|
H_{R}\right| \Psi _{m}\right\rangle \right| ^{2} \nonumber \\
\times \delta (\omega
-(E_{n}-E_{m})/\hbar),  \label{ramanme}
\end{eqnarray}
where $\mid \Psi _{n(m)}\rangle $ are eigenstates with energies
$E_{n(m)}.$ Such formula can be easily interpreted as a Fermi golden
rule averaged over the Boltzmann weight. 
To get informations on two-magnons scattering processes we should perform a
symmetry analysis of the matrix elements appearing in(\ref{ramanme}), and
discuss selection rules. Since the spin ladder Hamiltonian is invariant
under  translation along the legs, SU(2) rotation, and mirror along the
  leg direction, an eigenstate should be
characterized by a (lattice) momentum defined modulo $2\pi /a$ (where $a$ is
the lattice spacing), a spin and its parity under leg exchange. The Raman
operator defined in (\ref{hr}) is rotationally and translationally
invariant, and still invariant under leg exchange. As a result, the
selection rules impose that the states $\mid \Psi _{n}\rangle $ and $\mid \Psi
_{m}\rangle $ have the
same spin, momentum and parity under leg exchange. This implies, in
particular, that at $T=0$, transitions will only take place to states
of total momentum zero, spin zero and same parity as the ground
state. 
Let us now turn to concrete calculations.
 We consider the scattering
for $\overrightarrow{{E}_{I}}$ and
$\overrightarrow{{E}_{S}}$, parallel to the rung direction, thus we have

\begin{equation}
H_{R}=\frac{cste}{2}E_{I}E_{S}\sum_{i}\overrightarrow{{S}}_{1,i}\overrightarrow{{S}}_{2,i}
.
\label{hrr}
\end{equation}

In the following, we will evaluate the Raman intensity in the weak
coupling and in the strong coupling limit using the standard Matsubara
technique\cite{mahan_book} to calculate the correlator $\chi_R(\omega)$.

To evaluate the time ordered Raman susceptibility for the weakly coupled
chains, we will employ the Majorana fermion representation of the
spin-ladder Hamiltonian (\ref{hladder}) introduced by Shelton, Nersesyan and
Tsvelik in Ref. \onlinecite{shelton_spin_ladders}. 
The effective Hamiltonian is expressed in
terms of four interacting Majorana fermions. They comprise a
degenerate triplet $\xi _{\nu }^{a}(x)$ ($\nu =L$eft$,R$ight) with
bare mass $m_t=m=J_\perp$ and  a singlet, $\rho_{\nu }(x)$ with bare mass
$m_s=-3m$.
It has been argued in Ref. \onlinecite{shelton_spin_ladders} that the
effect of interactions was merely to renormalize the bare masses, so
that interactions could be neglected.
With this approximation, the spin ladder is described by the following
effective Hamiltonian:

\noindent

\[
H=\sum_{a=1,2,3} H_{m}[\xi ^{a}]+H_{-3m}[\rho ] 
\]

\noindent where

\begin{equation}
H_{\mu}[\zeta]=\left[ -i\frac{v_s}2\left\{ \zeta_R
\partial_x\zeta_R-\zeta_L\partial_x\zeta _L\right\}
-i\mu\zeta_L\zeta _R\right],
\label{hm}
\end{equation}

\noindent where $\mu$ stands for the triplet or singlet mass, and $\zeta$
is the corresponding triplet or singlet operator.
The thermal Green's function for the left and right moving triplet and
singlet Majorana fermions are defined by:

\begin{eqnarray}
G_{\mu \nu }^{t}(k,i\omega _{n}) &\equiv &\left\langle \xi _{\mu }^{\alpha
}(-\omega _{n},k)\xi _{\nu }^{\alpha }(\omega _{n},k)\right\rangle , \\
G_{\mu \nu }^{s}(k,i\omega _{n}) &\equiv &\left\langle \rho _{\mu }(-\omega _{n},k)\rho _{\nu }(\omega _{n},k)\right\rangle . 
\nonumber
\end{eqnarray}
whose explicit expressions are

\begin{eqnarray}
G_{RR}^{\alpha }(k,i\omega _{n}) &=&G_{LL}^{\alpha }(-k,i\omega _{n})=-\frac{%
i\omega _{n}+v_{s}k}{\omega _{n}^{2}+v_{s}^{2}k^{2}+m_{\alpha }^{2}}
\label{greenf} \\
G_{RL}^{\alpha }(k,i\omega _{n}) &=&G_{LR}^{\alpha }(k,i\omega _{n})^{\ast
}=-\frac{im_{\alpha }}{\omega _{n}^{2}+v_{s}^{2}k^{2}+m_{\alpha }^{2}} 
\nonumber
\end{eqnarray}
where $\alpha $ stands for $t$ (triplet) or $s$(singlet),  and $\omega
_{n}=(2n+1)\pi /\beta $ are the fermion Matsubara frequencies.  In terms 
of Majorana fermions, the Raman operator $\gamma
\sum_{i}\overrightarrow{{S}}_{1,i}\cdot \overrightarrow{{S}}_{2,i}$,
with $\gamma $ 
a constant, is expressed by 
\begin{equation}
\label{rop}
H_R=\gamma _{t}\overrightarrow{\xi_{R}}
\overrightarrow{\xi_{L}}+\gamma _{s}\rho_{R}\rho_{L},
\end{equation}
\noindent where $\gamma _{t}=m_{t}\gamma $
and $\gamma _{s}=m^{s}\gamma.$ To arrive at this expression, the marginal
term, already neglected in the derivation
of the Hamiltonian (\ref{hm}), has been discarded. Injecting this
expression into the
definition of the Raman susceptibility (\ref{ramansusc}) and applying Wick's
theorem, the time ordered expectation value at finite temperature can be
written as:

\begin{eqnarray}
\chi _{R}(i\omega _{n})=\frac 1 \beta \sum_{\nu_{n},\alpha }\gamma _{\alpha
}^{2}\int \frac{dq}{2\pi }[G_{RL}^{\alpha }(q
,i\omega _{n})G_{LR}^{\alpha }(-q,i(\omega
_{n}-\nu_{n})) \nonumber \\ 
-G_{RR}^{\alpha }(q,i\omega
_{n})G_{LL}^{\alpha }(-q,i(\omega _{n}-\nu_{n}))].
\label{ramansuscwc}
\end{eqnarray}

Explicitly, we have to compute the following integral and sum over the
Matsubara frequencies:

\begin{eqnarray}
\chi _{R}(i\omega _{n})=\frac 1 \beta\sum_{\nu_n,\alpha }\gamma _{\alpha
}^{2}\times \nonumber \\ 
\int \frac{dq}{2\pi }\left\{ \frac{m_{\alpha }^{2}-(i\nu_n+vq)(i(\omega
_{n}-\nu_n)+vq)}{(\nu_n^{2}+(vq)^{2}+m_{\alpha }^{2})((\omega
_{n}-\nu_n)^{2}+(vq)^{2}+m_{\alpha }^{2}}\right\} .  \label{ramansuscwcb}
\end{eqnarray}

In order to evaluate the Matsubara sum in (\ref{ramansuscwcb}), we have to
determine the residues of the four poles of the expression (\ref
{ramansuscwcb}) and multiply every residue with the value of the Fermi
function $n_{F}(z)=1/(\exp (\beta z)+1)$ at the pole. Adding the four terms
together yields

\begin{eqnarray}
\chi _{R}(i\omega _{n})=-\sum_{\alpha }\gamma _{\alpha }^{2}\int \frac{dq}{%
\varepsilon _{\alpha }(q)}(1-2n_{F}(\varepsilon _{\alpha }(q)))\times
\nonumber \\  2vq\left[ 
\frac{i\omega _{n}+2vq}{\omega _{n}^{2}+4\varepsilon _{\alpha }(q)^{2}}%
\right] ,
\end{eqnarray}
where we have introduced the notation $\varepsilon_\alpha (q)=\sqrt{%
(vq)^{2}+m_{\alpha }^{2}.}$ Thus performing the analytic continuation $(i\omega
_{n}\rightarrow \omega +i0_+ )$, taking the imaginary part and
performing the integral over $q$, we finally get: 

\begin{eqnarray}
{ \rm Im} \chi _{R}(\omega )=\pi \sum_{\alpha }\tanh \left( \frac{\omega }{4k_{B}T%
}\right) \gamma _{\alpha }^{2}\frac{\sqrt{\omega ^{2}-4m_{\alpha }^{2}}}{%
2\omega v}\nonumber \\ \times \Theta(|\omega|-2m_\alpha)  \label{wcres}
\end{eqnarray}

\begin{figure}
\centerline{\epsfig{file=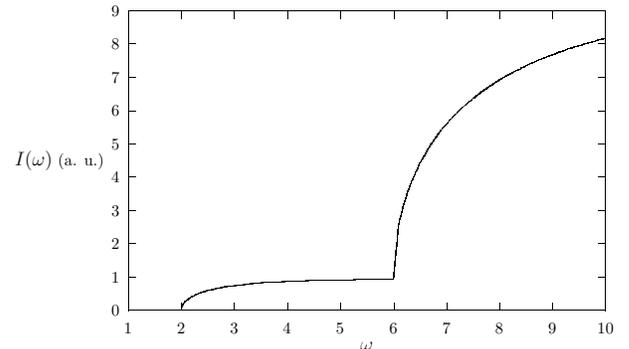,angle=0,width=8cm}}
\caption{Raman intensity in arbitrary units for $J_\perp \ll J$ at
$T=0 K$ obtained from the Majorana fermion approach. The
frequency $\omega$ is measured in units of the gap.}
\label{fig:majo}
\end{figure}

Formula (\ref{wcres}) implies the existence of a cusp singularity in the
Raman intensity at  twice the spin gap due to the triplet excitations 
and another singularity at six times the spin gap due to the singlet modes. 
As result, the noninteracting Majorana fermions
representation does not reproduce the Raman peak experimentally
observed\cite{srcuo_raman_spectra,cavo_raman_spectra,sugai_raman_sr14cu24o41}.
The spectra predicted by (\ref{wcres}) is plotted on
figure~\ref{fig:majo}. The absence of signal for $\omega$ smaller than
twice the gap is in qualitative agreement with numerical
simulations\cite{spinladder_raman_numerics}. 
It would be interesting to determine whether treating
properly  the
interactions between the Majorana Fermions can reproduce the
experimental peak at twice the spin gap. 
We now turn to a strong-coupling analysis of the Raman
susceptibility, using the bond operator
representation\cite{sachdev_bot} of quantum S=1/2
spins used by Gopalan, Rice and Sigrist\cite{gopalan_2ch} in
their mean field approach to spin ladders. In this representation, one
starts from weakly coupled rungs and introduces on each rung a singlet
$s^\dagger$ and three triplets $t^\dagger_\alpha$ ($\alpha=x,y,z$) boson
creation operators, that span the Hilbert space of a single rung when
acting on a vacuum state. Since the rung can be in either the singlet
or one of the triplet states, the condition :
\begin{equation}\label{eq:projection}
s^\dagger s +
\sum_\alpha t^\dagger_\alpha t_\alpha=1
\end{equation}
 has to be satisfied by the physical states.
The representation of the spins ${\bf S}_{1}$ and ${\bf S}_{2}$ in terms of
these singlet and triplet operators, is derived in
Ref.\cite{sachdev_bot,gopalan_2ch}. 
Substituting this operator representation of spins into the original
Hamiltonian, one ends up with an Hamiltonian quartic in boson fields.
Treating the singlet operator in a mean field approximation and
neglecting interactions among the triplets, one obtains the following
Hamiltonian quadratic in triplet operators\cite{gopalan_2ch}:

\begin{eqnarray}
H_{MF}=(\frac{J_{\bot }}{4}-\mu )\sum_{i,\alpha }t_{i,\alpha }^{^{\dagger
}}t_{i,\alpha }\nonumber \\
+\frac{Js^{2}}{2}\sum_{i,\alpha }(t_{i,\alpha }^{^{\dagger
}}+t_{i,\alpha })(t_{i+1,\alpha }^{^{\dagger }}+t_{i+1,\alpha }).
\label{meanfieldh}
\end{eqnarray}
The chemical potential term $\mu$ guarantees that the condtion
(\ref{eq:projection}) is satisfied on average. 
This Hamiltonian can be solved by Green's function method. One, first,
introduces the four Green's functions $G_{i,\alpha }(\tau )=-\langle T_{\tau
}t_{i,\alpha }(\tau )t_{0,\alpha }^{\dagger }(0)\rangle $, $\tilde{G}%
_{i,\alpha }(\tau )=-\langle T_{\tau }t_{i,\alpha }^{\dagger }(\tau
)t_{0,\alpha }(0)\rangle $, $F_{i,\alpha }(\tau )=-\langle T_{\tau
}t_{i,\alpha }(\tau )t_{0,\alpha }(0)\rangle $, $F_{i,\alpha }^{\dagger
}(\tau )=-\langle T_{\tau }t_{i,\alpha }^{\dagger }(\tau )t_{0,\alpha
}^{\dagger }(0)\rangle $ and their Fourier transforms. We have:
\begin{eqnarray}
G(k,i\omega _{n}) &=&-\left[ \tilde{G}(k,i\omega _{n})\right]
^{\ast }=\frac{%
i \omega _{n}+\Lambda _{k}}{\omega _{n}^{2}+\omega _{k}^{2}}
\label{eq:gf1} \\
F(k,i\omega _{n}) &=&F^{\dagger }(k,i\omega _{n})=\frac{2\Delta _{k}}{\omega
_{n}^{2}+\omega _{k}^{2}}  \nonumber
\end{eqnarray}
where $\nu _{n}=\frac{2n\pi }{\beta }$ and the following notation has been
introduced: $\omega_{k}^{2}=\Lambda_{k}^{2}-(2\Delta_{k})^{2},$ with $%
\Delta_{k}=Js^{2}/2\cos k$ and $\Lambda_{k}=J_{\perp }/4-\mu +Js^{2}\cos k$%
, recovering the dispersion  relation predicted by Gopalan, Rice and
Sigrist \cite{gopalan_2ch}. As shown in Ref.\cite{gopalan_2ch}, the parameters 
$\mu $ and $s$ are determined by solving the self-consistent saddle point
equations. Let us now turn to the calculation of the Raman intensity. 

\bigskip The Raman intensity is proportional to ${\rm Im}\chi
_{R}(i \omega _{n}\rightarrow \omega +i 0)$ where: 
\begin{equation}
\chi _{R}(i\omega _{n})=\sum_{\alpha ,\beta }\int_{0}^{\beta }d\tau
e^{i \omega_n}\langle
T_{\tau }(t_{\alpha }^{\dagger }t_{\alpha })(\tau )(t_{\beta }^{\dagger
}t_{\beta })(0)\rangle  \label{eq:definition}
\end{equation}

\begin{figure}
\centerline{\epsfig{file=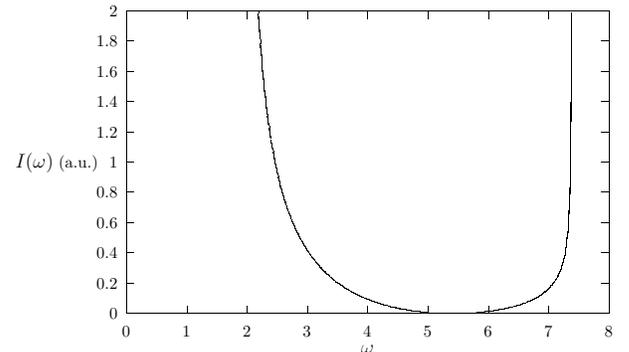,angle=0,width=8cm}}
\caption{Raman intensity in arbitrary units for $J_\perp/J=2$ at $T=0 K$. The
frequency $\omega$ is measured in units of the gap.}
\label{fig:intensity}
\end{figure}

By using the definition\ (\ref{eq:definition}) and applying Wick's theorem,
\ the following expression for the Raman susceptibility is obtained: 
\begin{eqnarray}
\chi _{R}(i \omega _{n})=\beta ^{-1}\sum_{\nu _{n}}\int \frac{dk}{2\pi }%
\left[ G(k,i\nu _{n})G(k,i\nu _{n}-i\omega
_{n})\right. \nonumber \\ \left. +F(k,i\nu _{n})F^{\dagger
}(k,i\omega _{n}-i\nu _{n})\right] .  \label{eq:raman_gf}
\end{eqnarray}
Performing the usual linear response calculation\cite{mahan_book},
 we obtain as a final result:

\begin{equation}
{\rm Im} \chi_{R}(\omega)=
\frac{\coth\left(\frac \omega {4 k_B
T}\right) \left[\left(\frac{ \omega}{2\left( J_\perp/4 -\mu
\right)}\right)^2-1\right]^2 }{4 \omega \sqrt{\left(\frac{2
J s^2}{ J_\perp/4 -\mu}\right)^2- \left[ \left(\frac{\omega}{2\left(J_\perp/4
-\mu\right)}\right)^2-1\right]^2}} 
\end{equation}
The Raman scattering spectra will show two peaks, one at energy $\omega
=2\omega_\pi=2\Delta _{s}$ corresponding to the bottom of the triplet
band, and a second one at $\omega
=2\omega_0$, corresponding to the top of the triplet band. Close to
the critical frequency $\omega^*$, $I(\omega)\sim
(\omega-\omega^*)^{-1/2}$. 
This bekavior can be easily understood by a density of states
argument. The resulting spectra is plotted in
figure~\ref{fig:intensity}. No signal is obtained for
$\omega<2\Delta_s$ in agreement with
numerics\cite{spinladder_raman_numerics}.  
Let us note that  in recent experiments, a
Raman scattering peak at twice the gap is observed 
 in $\mathrm{CaV_{2}O_{5}}$\cite{cavo_raman_spectra} 
where the spin-gap and the exchange constant are
estimated to be $\Delta _{s}\sim 400cm^{-1}$ and $J_{\perp }\sim
640K$. These results are in qualitative agreement with our theory. 
In the case of $\mathrm{Sr_{14}Cu_{24}O_{41}}$, the situation is more
complicated due to the coexistence in the structure
 of dimerized spin chains, having a
spin gap $\Delta_{\text{chain.}}=12 \mathrm{meV}$ \cite{tsuji_nmr} 
and of spin ladders having a spin gap $\Delta_{\text{ladder.}}\simeq
32$ meV\cite{eccleston_spin_dynamics_ladder}. 
In Ref. \onlinecite{srcuo_raman_spectra}, a peak was obtained
at $570 \mathrm{cm}^{-1} \simeq 71 \mathrm{meV} $ in Raman scattering
experiments on polycrystalline samples. Accordind to our
theory, this would lead to a spin gap of $\simeq 35 \mathrm{meV}$, in
agreement with neutron scattering datas. A more recent investigation
or Raman scattering on single crystals \cite{sugai_raman_sr14cu24o41}
identifies a peak at $498 \mathrm{cm}^{-1}$ as the Raman peak
associated with the gap. The peak at $569 \mathrm{cm}^{-1}$ is
identified with a $(0,0)$ gap. According to the authors of
Ref. \onlinecite{sugai_raman_sr14cu24o41}, the other peaks are
associated with bound states or single magnon light scattering.
It is known that bound states of magnetic excitations can be formed
below the gap in a spin ladder\cite{sushkov_ladder_boundstates,damle_ladder}.
In our treatment, we have been neglecting them altogether. They should
give rise to peaks below the threshold $2\Delta$, as has been observed
in  experiments \cite{sugai_raman_sr14cu24o41}. 
This problem is under investigation. 
To summarize, we have considered Raman scattering in a spin ladder
both in the weak coupling and the strong coupling approximation. We
have shown that only the strong coupling treatment gave rise to peaks
in the Raman intensity. Future directions include the consideration of
the effect of bound states on the Raman spectra. 
We thank T. Giamarchi and O. Parcollet for their remarks on the
manuscript. E. O. acknowledges discussion with K. Damle on the bound
states in a spin ladder.  E. O. acknowledges support from NSF under grant 
DMR 96-14999.

\bibliographystyle{prsty}

\bigskip

\end{document}